\documentclass[options,onecolumn]{mn2e}
\usepackage{epsfig}
\newif\ifAMStwofonts

\newcommand{\VEV}[1]{\langle#1\rangle}

\def\kms{\ {\rm Km\,s^{-1}}}

\def\hmpc{\ {\rm h^{-1}Mpc}}

 \title[The SZ effect from gas in the local universe]
{A full sky prediction of the SZ effect from diffuse hot gas in the local universe and the upper limit from the WMAP data}

\author[F. K. Hansen, E. Branchini, P. Mazzotta, P. Cabella, K. Dolag]
  {{F. K. Hansen$^1$, \thanks{E-mail: frodekh@astro.uio.no}}, {E. Branchini$^2$, \thanks{E-mail: branchin@fis.uniroma3.it}}, {P. Mazzotta $^{3,4}$, \thanks{E-mail: pasquale.mazzotta@roma2.infn.it}}, {P. Cabella$^{3}$, \thanks{E-mail: Paolo.Cabella@roma2.infn.it}},  {K. Dolag$^{5}$, \thanks{E-mail: kdolag@MPA-Garching.MPG.DE}}\\
$^1$ Institute of Theoretical Astrophysics, University of Oslo, P.O. Box 1029 Blindern, N-0315 Oslo, Norway\\
$^2$ Dipartimento di Fisica, Universit\`a di Roma TRE, Via della Vasca Navale 84, I-00146 Roma, Italy\\
$^3$ Dipartimento di Fisica, Universit\`a di Roma `Tor Vergata', Via della Ricerca Scientifica 1, I-00133 Roma, Italy\\
$^4$ Harvard-Smithsonian Center for Astrophysics, 60 Garden Street, Cambridge, MA02138, USA\\
$^5$ Max Planck Institut f\"ur Astrophysik, Karl-Scharzschild-Strasse 1, Postfach 1317, D-85741 Garching bei M\"unchen, Germany\\}

\pagerange{\pageref{firstpage}--\pageref{lastpage}}
\pubyear{2002}

\begin{document}

\label{firstpage}

\maketitle

\begin{abstract}
We use the PSC$z$ galaxy redshift catalogue combined with constrained simulations based on the IRAS 1.2 Jy galaxy density field to estimate the contribution of hot gas in the local universe to the SZ-effect on large scales. We produce a full sky Healpix map predicting the SZ-effect from clusters as well as diffuse hot gas within $80 \hmpc$. Performing cross-correlation tests between this map and the WMAP data in pixel, harmonic and wavelet space we can put an upper limit on the effect. We conclude that the SZ effect from diffuse gas in the local universe cannot be detected in current CMB data and is not a contaminating factor on large scales ($\ell<60$) in studies of the CMB angular anisotropies. However, for future high sensitivity experiments observing at a wider range of frequencies, the predicted large scale SZ effect could be of importance. 
\end{abstract}

\section{introduction}

The first high resolution full-sky observation of the cosmic microwave background (CMB) provided by the WMAP satellite has significantly improved estimates of the angular power spectrum and cosmological parameters (see Bennett et al. \shortcite{bennett} and references therein). It has also put stringent limits on the level of primordial non-Gaussianity and point source contribution in the CMB \cite{komatsu}. Also the hot gas in some nearby clusters was seen through the Sunyaev-Zeldovich (SZ) effect.

In this paper we investigate whether CMB observations, at the frequencies observed by the WMAP satellite, could be used to study the large scale properties of the local universe ($<80 \hmpc$) through the SZ effect of diffuse hot gas as well as nearby clusters and/or if this effect could  contaminate, to some level, the CMB at the largest scales. This is done through cross-correlation tests between an all-sky estimate of  the SZ thermal effect induced by the hot gas in the local universe and combinations of the WMAP observations at different frequencies. We also put limits on the contribution of the SZ effect in the local universe to the WMAP data. Uncertainties in the reconstructed positions of local cosmic  structures in the currently avaliable constrained simulations are significantly larger than 2 degrees, which corresponds to the smallest angular scale that we consider in our cross-correlation analysis. To circumvent this problem we adopt an alternative reconstruction  procedure that uses a further set of constraints provided by the observed positions of PSCz galaxies.

Recently there have been a number of papers focusing on cross-correlation tests to search for the SZ effect in the WMAP data; \cite{myers} found a detection of the SZ effect out to a $1^\circ$ angular distance of the cluster centres for clusters in the APM, ACO and 2MASS galaxy catalogues; \cite{carlos2} performed a cross-correlation test between 2MASS (out to $z\approx0.1$) and WMAP for small patches on the sky and found a strong detection; \cite{isw6} detected the thermal SZ effect at small angular scales by cross-correlating the WMAP data with the 2MASS galaxy catalogue; \cite{afshordisz} found an $8\sigma$ detection of the SZ effect for 116 X-ray clusters; \cite{isw2,isw5} reported a detection of the SZ effect of high redshift clusters by cross-correlating the WMAP data with large scale structure traced by the Sloan Digital Sky Survey (SDSS). All these detections of the SZ effect were found mainly on small angular scales and mostly due to high redshift clusters. On large angular scales, cross-correlations between WMAP observations and large scale structure have been found
\cite{isw1,isw2,isw3,isw4,isw5,isw6,iswwav,padm}. These correlations have been interpreted as a detection of the 
Integrated Sachs-Wolfe effect (ISW) and are not compatible with the signal expected from the SZ effect.
Some attempts to detect the SZ effect on large angular scales have also been performed. \cite{scaramella} used the 
distribution of Abell/ACO clusters to predict the SZ effect in the local universe and its detectability; \cite{diego} searched for cross-correlations between the WMAP data and the ROSAT diffuse X-ray background maps but found no correlations; \cite{carlos} used a set of optical and X-ray based catalogues and detected a cross-correlation due to the SZ effect at small scales but failed to detect an SZ signal caused by superclusters at larger scales; \cite{huffenberger} put limits on the power spectrum of the SZ effect by optimally combining cross power spectra between the different WMAP channels; limits have also been put on gravitational lensing as well as the SZ effect by cross correlating large scale structure traced by SDSS and the WMAP data \cite{hirata}. 

Here we focus on diffuse hot gas in the local low redshift universe ($z<0.05$) and perform cross-correlation tests of our predicted full sky SZ-map with the WMAP data on large angular scales ($>2^\circ$). A similar test was performed in \cite{banday} for the COBE data using a model for a predicted X-ray-emitting halo surrounding the Local Group, but no detection was found. First we test cross-correlations in pixel space for the full sky as well as for limited parts of the sky where the effect is expected to be largest. Then we perform a similar test in harmonic space for different multipole ranges. Finally, we use tests build on wavelets which have proven to be a very powerful tool for detecting cross-correlations \cite{iswwav}. Note that some of the clusters included in our analysis have been detected in some of the above mentioned papers, but only at small angular scales. Here we are interested only in the large angular scale effect introduced by the local hot gas.

The outline of the paper is as follows. In section \ref{sect:data} we describe the data sets used, the WMAP data and the PSC$z$ and IRAS galaxy catalogues and in section \ref{sect:sz} we describe how we obtain a prediction of the SZ effect from these data sets. Then in section \ref{sect:crosscor} we outline the cross-correlation tests applied in pixel-, harmonic- and wavelet space and present the results. Finally, these results are discussed in section \ref{sect:concl}.

\section{Description of the datasets}

\label{sect:data}

In this work, we have used the PSC$z$ galaxy catalogue combined with constrained simulations based on the IRAS 1.2Jy galaxy density field to infer the matter densities and hence the gas densities in the local universe.  From this we have made a prediction of the SZ effect for which we have been looking using the latest CMB data from the WMAP experiment. In this section, we summarise the properties of these three data sets.

\subsection{The WMAP data}

The WMAP experiment (Bennett et al. \shortcite{bennett} and references therein) observed the sky at 5 frequencies, 3 of which are interesting to studies of the CMB, the Q-band at $41 \textrm{Ghz}$, the V band at $61 \textrm{GHz}$ and the W band at $94 \textrm{Ghz}$. In this work, we have used the foreground cleaned V and W band maps (publicly available at the Lambda website\footnote{obtainable from the LAMBDA website: http://lambda.gsfc.nasa.gov/}) which have been co-added according to
 \begin{equation}
M^s=\frac{V+AW}{1+A},
\end{equation}
where $A$ is a constant. As the channel $Q$ may be more contaminated by foreground residuals than the other two channels at the scales considered in this analysis, we have chosen to exclude it from the analysis (as was done also by the WMAP team  e.g. Hinshaw et al. 2003 for the largest scales). We find that the value of $A$ optimising the ratio of the SZ effect to the variance of the CMB and noise is $A=0.59$. We also use the difference map $M^d=W-V$ which does not have a CMB component.  Note that we use the index $s$ for the optimal map and $d$ for the difference map. Furthermore, the galactic cut and point source mask $Kp2$ (which we extend by 2 degrees along the rim of the cut) used by the WMAP team (and available at the Lambda website) have been applied to avoid possible sources of contamination. Note that the unobserved region of the PSC$z$ catalogue is already contained within this mask. The maps are all in the Healpix pixelisation \cite{healpix} with resolution parameter $N_{side}=512$ corresponding to pixels with edges of about $7'$.

\subsection{The PSC$z$ catalogue and the galaxy density field}

\label{sect:pscz}

The PSC$z$ redshift catalogue lists the angular positions, redshifts and fluxes of $\sim 15,500$ IRAS PSC galaxies selected with a flux at 60 $\mu m $ larger than $0.6$ Jy. A detailed description of the selection criteria, star-galaxy separation algorithm and the procedures adopted to exclude galactic cirrus, are given in \cite{saunders} to which we refer the interested reader. For our purposes the most interesting features of the IRAS PSC$z$ catalogue is its very large sky coverage ($\sim 84 \%$ of the sky, overlapping very well with the part of the sky outside the $Kp2$ sky cut used by WMAP) and depth (the median redshift is $cz \sim 8500 \kms$). In this work  we consider a subsample of galaxies within a radius of $80 \hmpc$ for which distances have been obtained from redshift using the iterative procedure of \cite{branchini}. The flux-limited nature of the catalogue causes the number of objects to decrease with distance. The average galaxy-galaxy separation of the sample has been computed by \cite{branchini}

\begin{equation}
 \langle l(r) \rangle = \left\{\begin{array}{ll}
   \langle l_0 \rangle& {\rm if} \ \ r\le 6 \hmpc  \\
  A\langle l_0 \rangle
\left(r\right)^{-0.36}\left(1+ {{r}^2\over{r_{\star}^2}} \right)^{0.61}
& {\rm if } \ \ r>6 \hmpc
\end{array}\right.
\label{eq:self}
\end{equation}

where $r$ is the proper distance in $\hmpc$, $r_{\star}=87 \hmpc$,  $\langle l_0 \rangle=2.17 \hmpc$ and the parameter $A$ guarantees continuity at $r=6 \hmpc$. To obtain a continuous density field from galaxy positions with constant sampling errors throughout the sample, we have divided our spherical volume into spherical shells $10 \hmpc$ thick and have smoothed the galaxy density field using a set of Gaussian filters of increasing radius. The latter have been fixed by imposing a constant angular smoothing of about $2^{\circ}$, as required by our cross correlation analysis. As a consequence the galaxy distribution within the spherical shells with external radii of 20, 30, 40, 50, 60, 70 and 80 $\hmpc$ have been interpolated on a cubic grid and smoothed with Gaussian filters of radii (FWHM) 0.7, 1.05, 1.4, 1.75 2.1, 2.45 and 2.8 $ \hmpc$. Eq.~\ref{eq:self} shows that this filtering procedure guarantees at least one object per resolution element, and thus keeps the shot noise errors at an acceptable level.

\subsection{Constrained simulation}

\label{sect:simul}

In this work we also use a third dataset generated by the cosmological hydrodynamic simulation of our local universe performed by \cite{dolag}. The simulation is based on a previous numerical experiment performed by \cite{mathis} aimed at reproducing the mass distribution within 12,000 $\kms$ traced by the IRAS galaxies in the 1.2Jy redshift survey \cite{fisher}. 
The original galaxy distribution was smoothed with a Gaussian filter of radius $5 \hmpc$ and traced back in time at $z=50$ using the method of \cite{kolatt}. This initial field was used as Gaussian constraint for an otherwise random realisation of a flat $\Lambda_{CDM}$ cosmology with $\Omega_m=0.3$, Hubble constant of $H_0=70 \ {\rm Km\,s^{-1}Mpc^{-1}} $ and a rms density fluctuation $\sigma_8=0.9$. The initial density field, specified in a box of $240 \hmpc$ with a high resolution region covering a sphere of 80 $\hmpc$, has been evolved forward in time using the latest version of GADGET code \cite{springel} including gas-physics with smoothed particle hydrodynamics in the high resolution region. As shown by \cite{mathis} the simulated mass distribution at $z=0$ reproduces well the main characteristics of the most prominent nearby cosmic structures, including clusters like Virgo and Coma, the Perseus-Pisces and the Great Attractor complexes. The hydrodynamic re-simulation of these initial conditions has been used to study the propagation of cosmic rays through the local universe \cite{dolag}. In this work, we have considered the simulated mass and gas distribution at $z=0$ within the high resolution spherical region of radius 80 $\hmpc$  and applied the same procedure adopted for the PSC$z$ datasets, with the purpose of smoothing the mass density and gas density and temperature fields on an angular smoothing scale of $2^{\circ}$. To minimise  boundary effects in the smoothing procedure we have assigned gas density and temperature to dark matter density also in each resolution element beyond 80 $\hmpc$ according to the gas vs. mass relations  measured in the high resolution region containing the simulated gas particles. The end product consist of values of the dark matter density, gas density  and temperature specified at the same locations as the PSC$z$ galaxy density field, within 
 the same set of spherical shells and smoothed with the same Gaussian filters specified 
in Section \ref{sect:pscz}.

\section{Method to obtain the SZ effect from the galaxy catalogue}

\label{sect:sz}

In this section, we describe how we use the galaxy catalogue to derive an estimate of the gas density and temperature in the local universe. From this estimate, we make a spherical projection to make a prediction for the SZ effect. The projection is made onto a Healpix-pixelised sphere with resolution parameter $N_{side}=64$ corresponding to pixel edges of roughly $0.9^\circ$.

\subsection{Mapping the local hot gas}

\label{sect:maps}

The goal of the smoothing procedure specified in the previous sections is to obtain a (Gaussian FWHM) $2^{\circ}$ resolution map of the predicted, cumulative SZ effect generated by all cosmic structures within 80 $\hmpc$.  Since the SZ distortions are fully specified by the  density and temperature of the gas along the line of sight, one may think that reliable SZ maps can be obtained directly from the constrained hydrodynamical simulations. This is certainly the case for most statistical analyses but for cross-correlation studies like the one we wish to perform here. Indeed, the constraints in the simulations are only effective above a (Gaussian) resolution scale of 5 $\hmpc$, corresponding to an angular scale $>3.5^{\circ}$ within our spherical volume. Since the actual positions of cosmic structures within each resolution  element are essentially random, we cannot directly cross-correlate the SZ maps obtained from the hydrodynamical simulation with the temperature fluctuations of the CMB with the desired angular resolution of $2^{\circ}$. The PSC$z$ galaxy density maps described in section \ref{sect:pscz} can be used to circumvent this problem. The mass overdensity field, $\delta_m$ in the constrained simulation has been obtained from the IRAS 1.2Jy galaxy density field assuming that IRAS galaxies trace the underlying mass distribution. On the other hand \cite{teodoro} have shown that the IRAS 1.2Jy and PSC$z$ galaxy density fields are in very good agreement (apart from a monopole mismatch that does not affect our analysis) and thus that a tight relation exists between $\delta_\mathrm{PSCz}$ and $\delta_m$. We can therefore specify the gas properties at the correct spatial locations by rank ordering the overdensity field  $\delta_{PSCz}$ and $\delta_m$ while keeping the original spatial association between $\delta_m$ and the gas density and temperature in the simulation. 
As a result, we are able to specify 
the temperature $T$, and the electron density $n_e$ in the gas
associated to the cosmic structures traced by PSC$z$ galaxies, hence allowing a cross 
correlation analysis with  a resolution of $2^{\circ}$.

\subsection{The SZ effect from line of sight integrals}

The  Sunyaev \& Zel'dovich effect \cite{sz0} arises due to the inverse Compton scatter of CMB photons against the hot and diffuse  electron gas trapped in the potential well of the cosmic web and of each cluster of galaxies. In the latter case these electrons are also responsible for their
X-ray emission.  The CMB temperature  change across a single line of sight is given by:
\begin{equation}
{\Delta T_{\rm CMB} \over T_{\rm CMB}} =  y_c s(x)
\end{equation}
 where $x=h\nu/kT_{\rm CMB}$. The Comptonization parameter is
\begin{equation}
\label{eq:y}
y_c = \int {kT  \over mc^2} n_e \sigma _T dl,
\end{equation}
where $n_e$ and $T $ are the electron density and temperature respectively, $\sigma _T$  is the Thomson cross section, and the integral is over a line of sight. The spectral form of this ``thermal effect" is described by the function $s(x) = [x\cdot \coth(x/2)-4]$,  which is negative (positive) at values of $x$ smaller (larger) than $x_0=3.83$, corresponding to a critical frequency  $\nu_0=217 \textrm{GHz}$. At the frequencies considered here, the temperature variation is always negative with values $s(x)=-1.55$ in the W channel, $s(x)=-1.80$ in the V channel and $s(x)=-1.91$ in the Q channel.

To obtain a map of the SZ effect, we need to integrate equation (\ref{eq:y}) along the line of site in all the shells described in section \ref{sect:data}. Each of the shells are pixelised in a 3D grid of $128\times128\times128$ pixels. Before making the line of sight integral, we made a 3D linear interpolation to obtain the shells on a $1280\times1280\times1280$ cubic grid. From this, we performed the following integration:
\begin{equation}
y_i=\frac{1}{\Delta\Omega}\frac{k\sigma_T}{mc^2}\sum_{j\epsilon i}T^jn_e^j\delta\Omega_j\delta l_j,
\end{equation}
where $i$ is pixel number on the Healpix grid, $j$ is grid element in the interpolated 3D grid and the sum is performed over all elements $j$ contained in the solid angle spanned by Healpix pixel $i$. The temperature and electron density at gridpoint $j$
are given by $T^j$ and $n_e^j$ (obtained from the procedure described in section \ref{sect:maps}), $\delta\Omega_j$ and $\delta l_j$ are the mean solid angles and distances for each gridpoint $j$ and finally $\Delta\Omega=4\pi/(12N_{side}^2)$
is the solid angle of each Healpix pixel. 
As all shells were smoothed with a 3D filter corresponding to Gaussian angular beam of FWHM $2^\circ$, our final map has this resolution. 

\begin{figure}
\begin{center}
\leavevmode
\epsfig {file=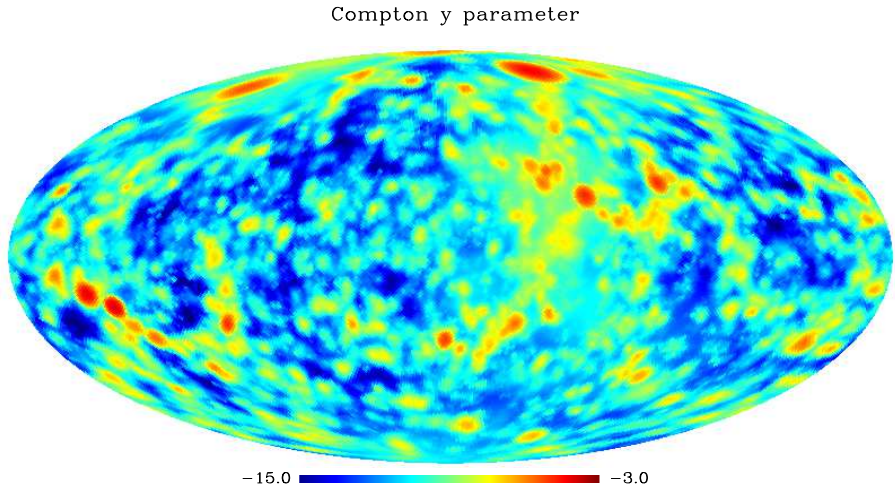,height=12cm,width=7cm,angle=90}
\caption{Our prediction of the Compton y parameter from the local SZ effect (the colour scales indicate $\log[y]$). The plot is in galactic coordinates with galactic longitude $l=0$ in the middle and increasing leftwards.}
\label{fig:SZmap}
\end{center}
\end{figure}

\begin{figure}
\begin{center}
\leavevmode
\epsfig {file=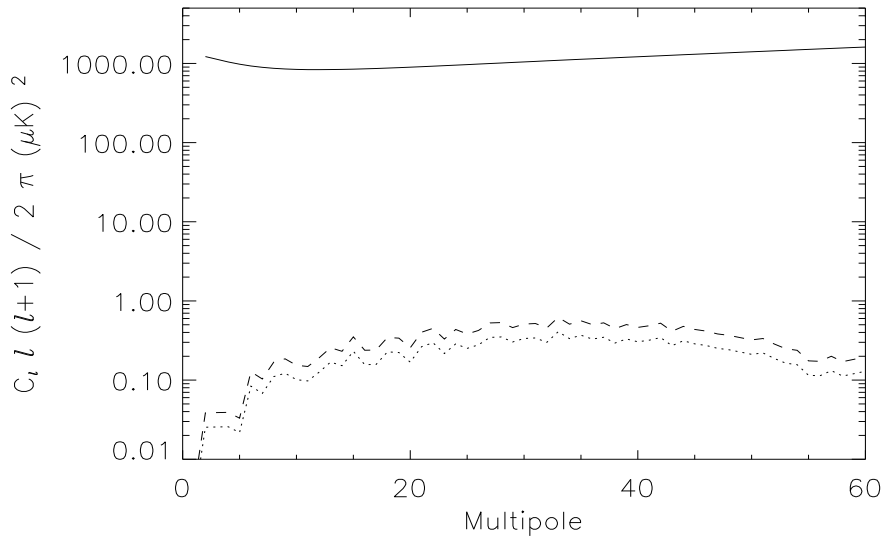,height=8cm,width=12cm}
\caption{The power spectrum of the local SZ-effect. The solid line shows the CMB power spectrum with the best fit WMAP parameters. The dashed and dotted lines show the power spectrum of the predicted local SZ-effect at the WMAP Q ($41\textrm{GHz}$) and W ($94\textrm{GHz}$) frequencies respectively. }
\label{fig:SZcl}
\end{center}
\end{figure}

In figure \ref{fig:SZmap} we show the resulting map of the Compton y parameter. We can see that the most dominant structure is the Virgo cluster on the upper right side. We have also plotted the power spectrum of this map at two different frequencies in figure \ref{fig:SZcl}. The dashed line corresponds to the Q ($41\textrm{GHz}$) frequency and the dotted line to the W ($94\textrm{GHz}$) frequency. Because of the very limited redshift range considered, the power spectrum starts to fall of at the smallest scales. We see that the predicted effect is considerably lower than the CMB power spectrum (solid line) and could not be distinguished from the cosmic variance at these scales.

\section{Upper limits on the SZ effect from the WMAP data: Cross correlations in pixel, harmonic and wavelet space}

\label{sect:crosscor}

Having obtained a map of the predicted SZ effect from gas in the local universe, the next step is to make a cross-correlation test between CMB data and the SZ map. We perform this cross-correlation in 
pixel, harmonic and wavelet space, using the two less foreground contaminated WMAP channels, V and W. We apply the test both on the optimally combined map which is dominated by the CMB and on the difference map in which the CMB disappears and only components which depend on the frequency remain.

\subsection{Estimating the SZ contribution in the WMAP data: Method}

Before cross-correlating the WMAP data with the SZ prediction obtained above, we must smooth both maps to a common resolution. As described above, the map of the SZ effect has been smoothed with a Gaussian beam of $2^\circ$ FWHM. Due to the limited resolution of the PSC$z$ map, this is also the highest resolution which can be used in the cross-correlation procedure. For the CMB data, we have deconvolved the map with the beam of the experiment (being of the order $14'$ FWHM) and convolved it with a $2^\circ$ FWHM Gaussian beam. Before performing the deconvolution/convolution operation, the map was multiplied with the $Kp2$ galaxy and point source mask, and a refilling procedure, described in \cite{curvat,cabella}, was applied. In the refilling procedure, the parts of the map just outside the point source holes where mirrored into the holes. The galactic cut was filled mirroring the data north and south of the cut. After the convolution, the map was degraded to Healpix resolution 
 $N_{side}=64$ which is the resolution of the SZ map and the $Kp2$ galactic cut extended with 2 degrees along the rim (accounting for the inaccuracies introduced by the smoothing inside the galactic cut) was applied.

We now describe the cross-correlation tests that we performed in the three different spaces, pixel space, harmonic space and wavelet space. The pixel space approach has the advantage that different regions of the sky can be tested for a possible detection whereas the harmonic space method can test cross-correlations at different multipoles. Finally, the wavelet test can check for correlations at different wavelet scales.

\subsubsection{Cross correlation in pixel space}

\label{sect:pix}

We model the CMB map obtained by combining optimally the the V and W channels (see Section \ref{sect:data}) as the sum of three components, CMB, noise, and SZ:
\begin{equation}
M^s_i=m_i^{CMB}+n_i^s+Cm_i^{SZ},
\end{equation}
where $i$ is pixel number on the Healpix grid, $m_i^{CMB}$ is the pure CMB
contribution, $n_i^s$ is the noise and $m_i^{SZ}$ is the predicted SZ effect
normalised in such a way that $C=1$ if the prediction is correct. Our aim
is to estimate $C$ from the WMAP data.
Consequently the difference map is modelled as:
\begin{equation}
M^d_i=n_i^d+Cm_i^{SZ},
\end{equation}
where again $m_i^{SZ}$ is normalised so that $C=1$ is the expected value of $C$.
We use a $\chi^2$ approach to estimate C;
\begin{equation}
\label{eq:chi2}
\chi^2=(\mathbf{m}-C\mathbf{m}^{SZ})^T\mathbf{C}^{-1}(\mathbf{m}-C\mathbf{m}^{SZ})
\end{equation}
which is minimised for the best estimate $\hat C$;
\begin{equation}
\label{eq:cest}
\hat C=\frac{\mathbf{m}\mathbf{C}^{-1}\mathbf{m}^{SZ}}{\mathbf{m}^{SZ}\mathbf{C}^{-1}\mathbf{m}^{SZ}}.
\end{equation} 
The correlation matrix $\mathbf{C}$ is given by $C_{ij}^s=\VEV{(m_i^{CMB}+n_i^s)(m_j^{CMB}+n_j^s)}$ for the optimal combination map and $C_{ij}^d=\VEV{n_i^dn_j^d}$ for the difference map. Different sets of pixels will be included in the vector $\mathbf{m}$ and the correlation matrix $\mathbf{C}$ in order to test both the full sky as well as only the areas where the SZ effect is expected to be largest. It is important to say that the map resolution $N_{side}=64$, corresponding to a pixel size of $0.9^\circ$, is much smaller that the smoothing length of $2^\circ$ used to generate the SZ maps (see section \ref{sect:maps}). This creates singularities in the correlation matrix. Thus for this case only, we degrade the maps used for the cross-correlation test to the Healpix resolution $N_{side}=32$ (that correspond to a pixel size of $1.8^\circ$). We neglected the contribution from the noise correlation matrix to the total correlation matrix $\mathbf{C}$ for the optimal combination map as this allowed us to calculate its elements fast using the theoretical formula for the CMB two point correlation function. Comparing to the total correlation matrix including noise obtained from 1000 Monte-Carlo simulations, we found that the error was at most a few percent. For the difference map, we found from Monte-Carlo simulations that the pixel-pixel correlations in the noise introduced by the smoothing was limited to the 2-3 neighbouring pixels which justifies using the diagonal approximation.

In order to test the variance of $\hat C$ on maps with no SZ effect, we produced 1000 simulated maps. We estimated $\hat C$ on each of these maps applying the following procedure:
\begin{itemize}
\item We generated 1000 realisations of the CMB from the best fit WMAP power spectrum, convolving with a beam and adding noise according to the specifications of the WMAP experiment.
\item We multiplied the map with the $Kp2$ mask and performed the refilling procedure described above for the data.
\item We deconvolved with the beam and convolved with the $2^\circ$ FWHM Gaussian beam, degrading the map to resolution parameter $N_{side}=32$ and multiplied with an extended galactic cut, following the same procedure as for the data.
\item We used this map to estimate $\hat C$ according to equation (\ref{eq:cest}).
\end{itemize}

\subsubsection{Cross correlation in harmonic space}
 
In this second analysis we
apply a spherical harmonic transform to the data and SZ map, obtaining a set of coefficients $a_{\ell m}$
\begin{equation}
a_{\ell m}=\sum_iM_iY_{\ell m}^i,
\end{equation}
where $M_i$ is the map, $Y_{\ell m}$ are the spherical harmonic functions, all taken at pixel $i$. The sum is performed over all pixels $i$ outside the mask given by the extended $Kp2$ cut at resolution $N_{side}=64$. 
We apply the procedure described in the previous section with the $a_{\ell m}$ as the elements of the vectors $\mathbf{m}$ and the correlation matrix now beeing given by $C_{\ell m,\ell' m'}=\VEV{a_{\ell m}a_{\ell'm'}}$. 
Here we use the diagonal approximation of the correlation matrix. To test the correlations at different scales, 
we performed the analysis in a set of different multipoles ranges.

\subsubsection{Cross correlation in wavelet space}

\label{sect:wavmeth}

The ability of the wavelets to amplify features at particular scales makes them a very powerful tool for CMB analysis. In this context, they have been used for denoising \cite{sanz}, extracting point sources \cite{cayon,tenorio,wavpoint}, detecting non-Gaussianity \cite{cabella,vielva,cruz,wavng,barreiro,forni,starck}, estimating the non-linear coupling constant $f_{NL}$ \cite{wavng2,wang,cabellab} and to put constraints on the topology of the universe \cite{wavtopo}. 
Furthermore, the strongest detection of the ISW effect was found using a cross-correlation test in wavelet space \cite{iswwav}, showing the power of wavelets for cross-correlation analysis.

The wavelet coefficients of a spherical function $T(\theta,\phi)$ can be defined as,
\begin{equation}
w(R,\theta,\phi)=\int d\Omega' \psi(\Delta\theta',R)T(\theta',\phi'),
\end{equation}
where the integration is performed over the whole sphere in $(\theta',\phi')$, $\Delta\theta'$ is the angular distance between the points $(\theta,\phi)$ and $(\theta',\phi')$, $R$ is the wavelet scale and $\psi$ is the Spherical Mexican Hat Wavelet given by \cite{wavng}
\begin{equation}
\psi(\Delta\theta,R)=\frac{1}{\sqrt{2\pi}N(R)}\left[1+\left(\frac{y}{2}\right)^2\right]^2\left[2-\left(\frac{y}{R}\right)^2\right]e^{-y^2/(2R^2)},
\end{equation}
where $y=2\tan(\Delta\theta/2)$ and $N(R)=R\sqrt{1+R^2/2+R^4/4}$.
The starting point for the wavelet analysis is as above, the convolved and degraded CMB map. We construct the cross-correlation coefficients defined as
\begin{equation}
\label{eq:cr}
c_R=\sum_i w_i(R)w_i^{SZ}(R)
\end{equation}
where $w_i$ and $w_i^{SZ}$ are the wavelet coefficient in pixel $i$ and scale $R$ for the observed (simulated) map and the SZ map respectively. The sum is performed over all pixels 
outside the wavelet mask at scale $R$ defined as the Kp2 mask (without point sources) extended with 2 degrees as well as the angle $2.5R$ (to account for inaccuracies introduced by the convolution inside the galactic cut, see Vielva et al. 2004a) around all edges.

As for the previous tests, we aim at estimating the constant $C$ 
by minimising a similar $\chi^2$ function to obtain the following best estimate $\hat C$
\begin{equation}
\hat C=\frac{\mathbf{c}\mathbf{C}^{-1}\mathbf{c}^{SZ}}{\mathbf{c}^{SZ}\mathbf{C}^{-1}\mathbf{c}^{SZ}},
\end{equation}
where the elements of the vector $\mathbf{c}$ are the coefficients $c_R$ of equation \ref{eq:cr}, the elements of the vector $\mathbf{c}^{SZ}$ are given by $c_R^{SZ}=\sum_i (w_i^{SZ}(R))^2$ and the scale-scale correlation matrix $C_{RR'}=\VEV{c_Rc_{R'}}$ 
is obtained from 1000 Monte-Carlo simulations. 
We also estimate $\hat C$ independently at each scale $R$ to detect possible correlations limited to certain scales.

\subsection{Estimating the SZ contribution in the WMAP data: Results}

In this section, we show the results of the cross-correlation tests outlined above.

\subsubsection{Cross correlation in pixel space}

\begin{figure}
\begin{center}
\leavevmode
\epsfig {file=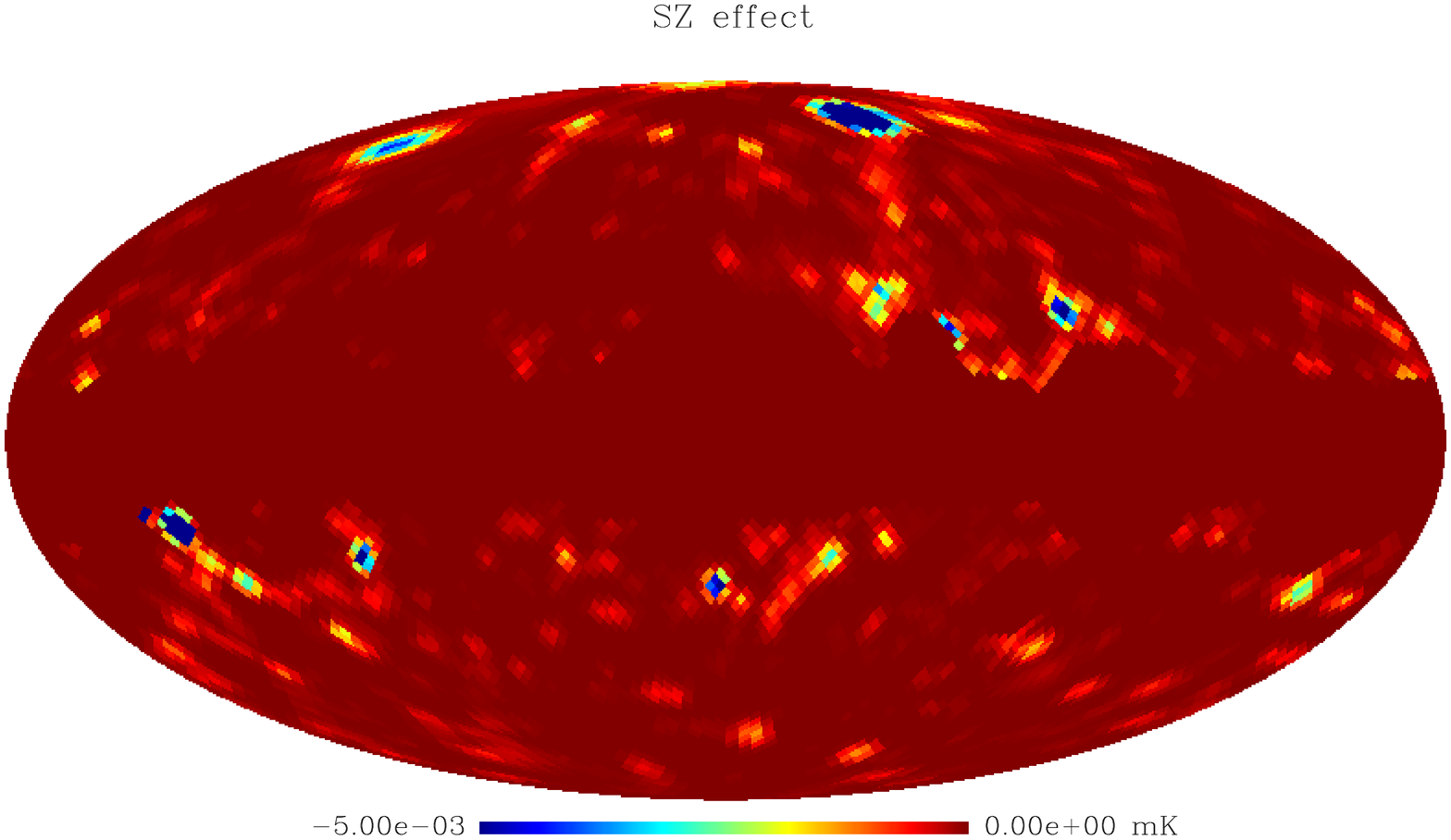,height=12cm,width=7cm,angle=90}
\epsfig {file=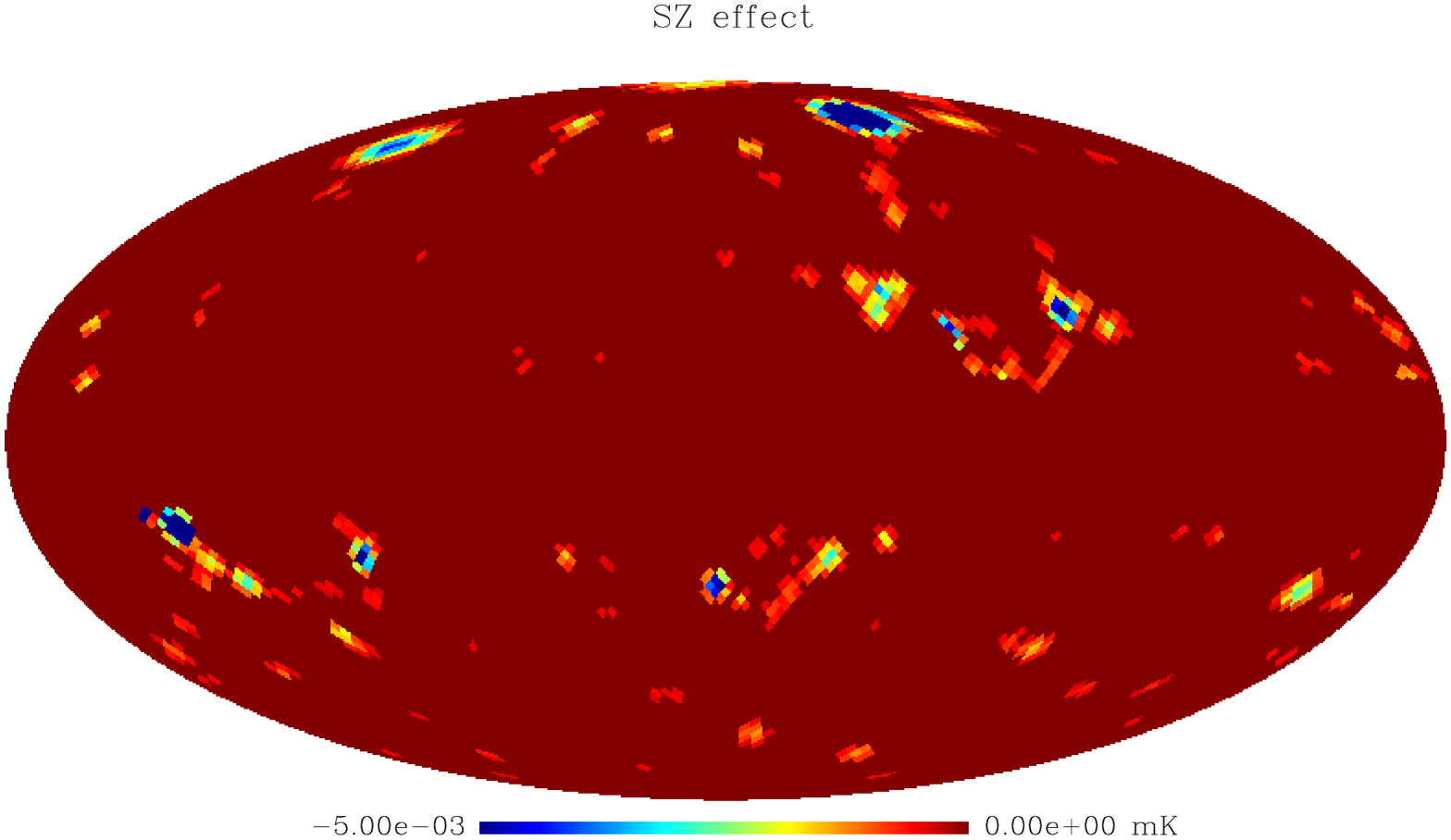,height=12cm,width=7cm,angle=90}
\epsfig {file=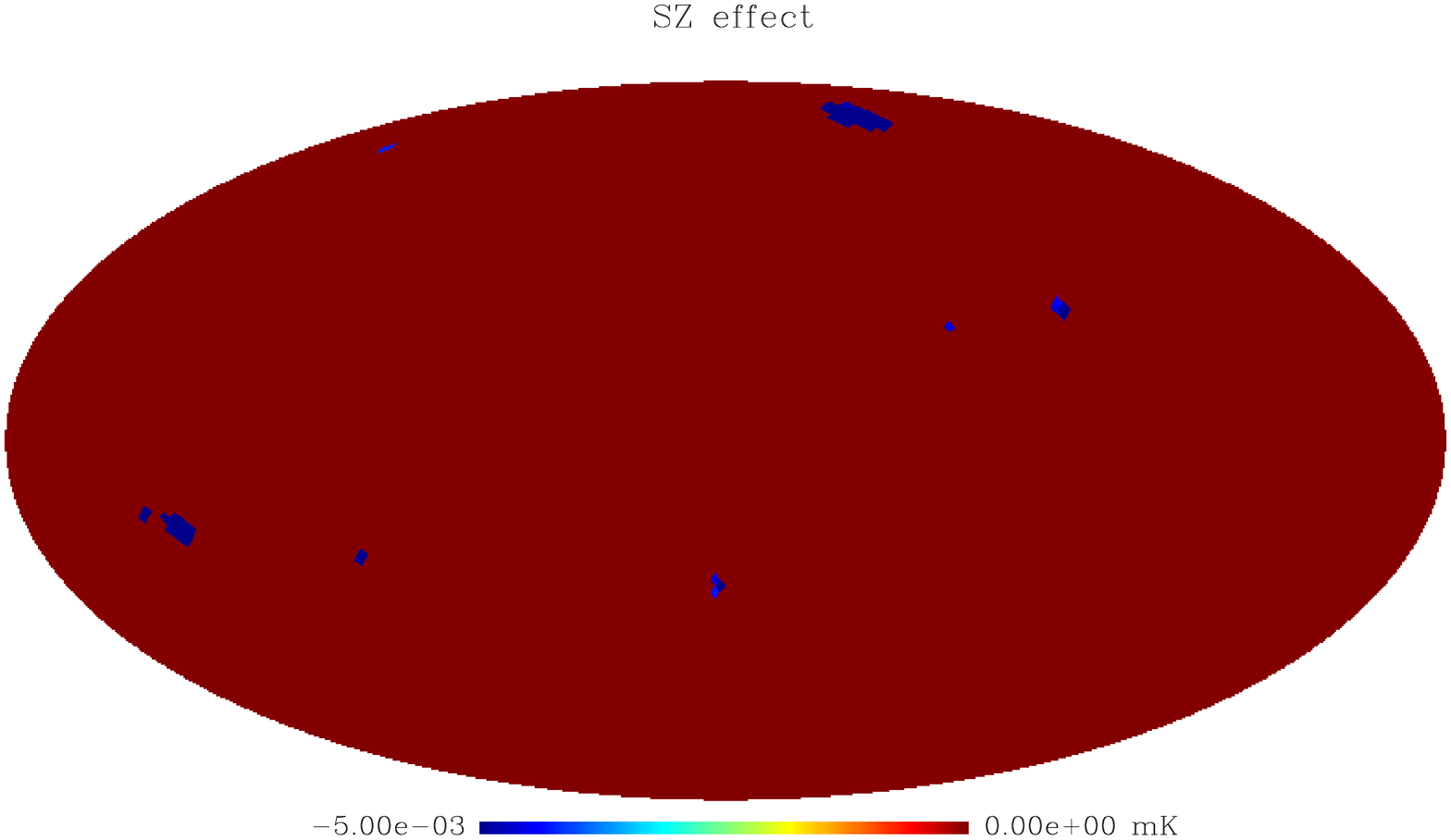,height=12cm,width=7cm,angle=90}
\caption{The masks used in the pixel space analysis. The mask (a) shown in the upper figure excludes only the pixels inside 
the extended $Kp2$ cut. The mask (b) shown in the middle excludes in addition all pixels where the expected Compton parameter $y<10^{-7}$ and finally the mask (c) shown in the lower figure excludes all pixels with $y<10^{-6}$.}
\label{fig:SZmasks}
\end{center}
\end{figure}

\begin{center}
\begin{table}
\caption{Estimates of $\hat C$ in pixel space with $1\sigma$ errors \label{tab:pixel}}
\begin{tabular}{|c|c|c|c|c|c|c|c|c|c|c|c|c|c|c|}
\hline
mask  & a & b & c  \\
combined map & $0.40\pm1.03$ & $0.36\pm1.12$ & $-0.45\pm1.60$ \\
difference map & $-1.00\pm1.44$ & $0.00\pm1.44$ & $-1.69\pm1.69$ \\
combining the two estimates & $-0.07\pm0.85$ & $0.23\pm0.93$ & $-0.46\pm1.66$ \\
\hline
\end{tabular}
\end{table}
\end{center}

The pixel space analysis has been carried out on three different sets of pixels, (a) all pixels outside the extended Kp2 cut, (b) all pixels outside the extended Kp2 cut where the expected Compton parameter $y>10^{-7}$ and (c) all pixels outside the extended Kp2 cut where the expected Compton parameter $y>10^{-6}$. The predicted SZ effect in the W channel outside these masks is shown in figure \ref{fig:SZmasks}. In table  \ref{tab:pixel} we show the resulting $\hat C$ for each of these sets of the optimal map and the difference map. We find that the error bars are too large to say anything on a clear detection of the expected SZ effect from hot gas in the local universe.

As the error bars of $\hat C$ on the combined map mainly comes from the variance of the CMB whereas the error bars of $\hat C$ on the difference map only comes from instrumental noise, we would expect these two estimates to be rather uncorrelated. In fact, Monte-Carlo simulations show that they are correlated at less than the $3\%$ level which means that combining the two statistics could yield smaller error bars on the total estimate of $\hat C$. The lower row in table \ref{tab:pixel} shows the estimate $\hat C$ found by combining the two statistics (weighting with variances and covariances). The strongest result in this paper is found for the smallest mask (a), we find $C=-0.07\pm0.85$ at the $1\sigma$ level, that is, we find an upper limit of $C<1.63$ at the $2\sigma$ level. Thus, at the $2\sigma$ level, the local SZ effect cannot exceed the prediction by more than $63\%$.

\subsubsection{Cross correlation in harmonic space}

\begin{figure}
\begin{center}
\leavevmode
\epsfig {file=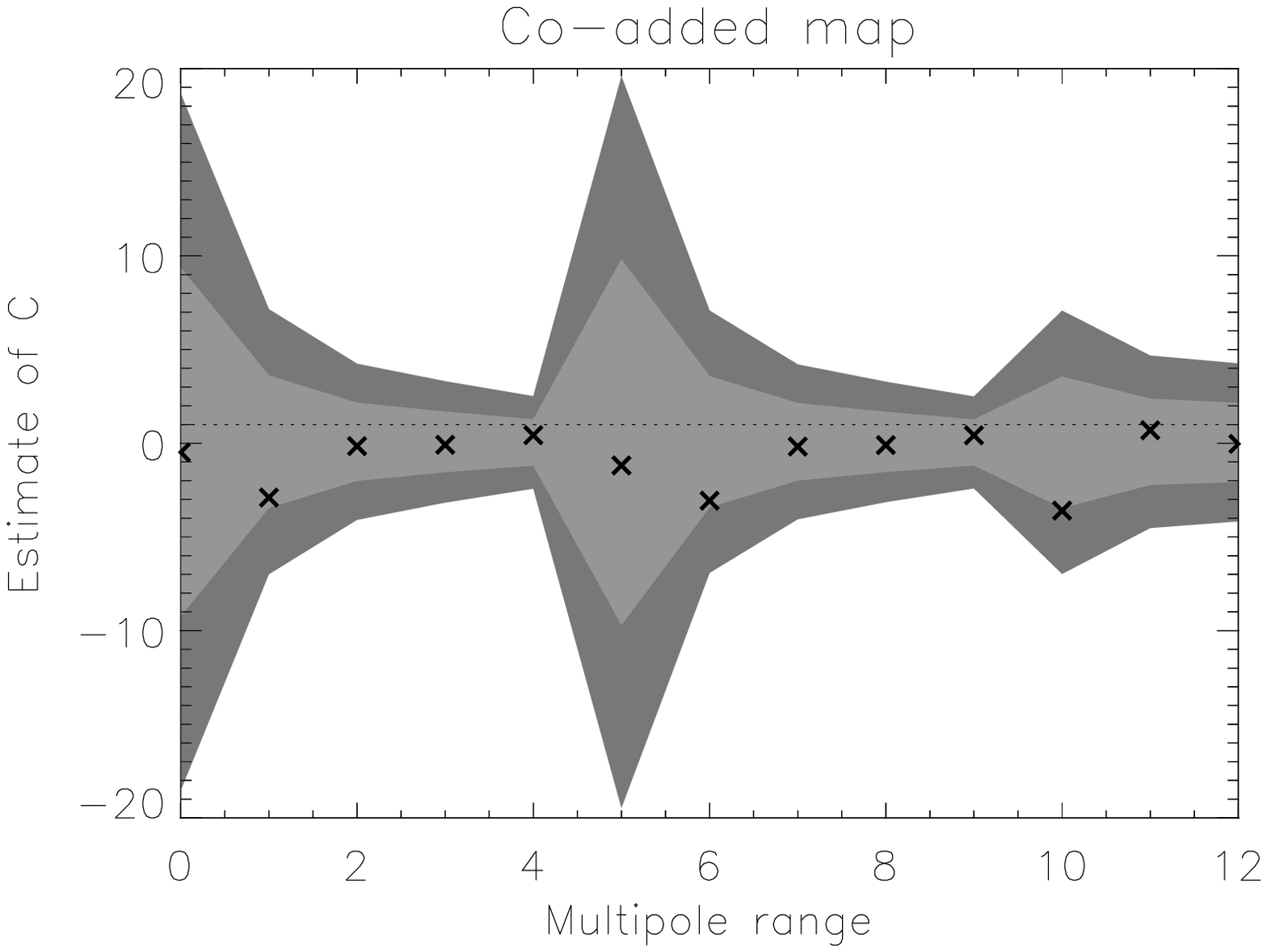,height=8cm,width=8cm}
\epsfig {file=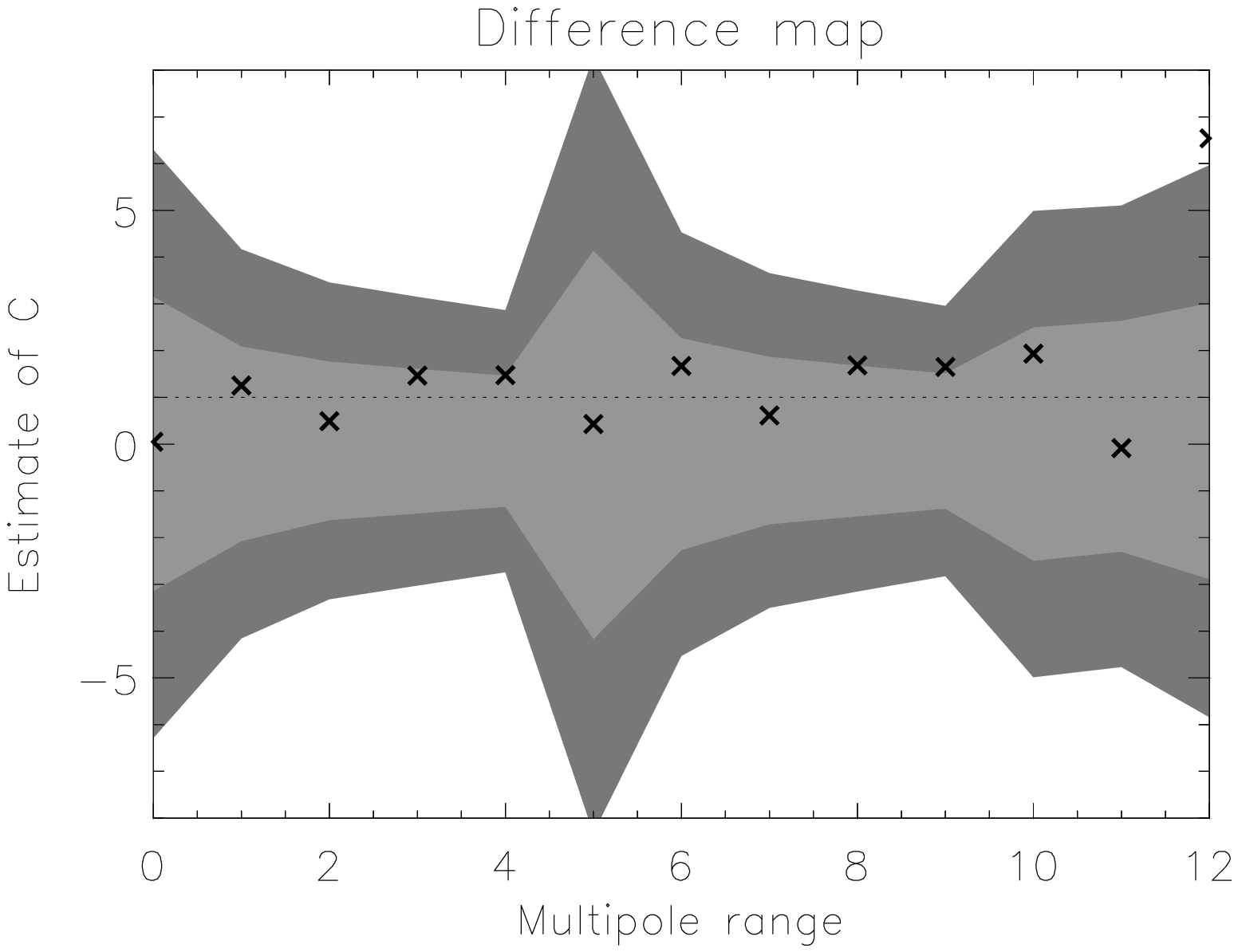,height=8cm,width=8cm}
\caption{The estimate of $C$ in harmonic space using the co-added V+W map (left plot) and the difference map (right plot). The grey bands show the 1 and 2 $\sigma$ level set from 1000 Monte-Carlo simulations of pure CMB maps, the dotted line shows the average of 1000 simulated CMB maps with the expected level of SZ-effect added. The crosses show the result from cross-correlation with the WMAP data. The multipole ranges corresponding to the abscissa values are given in table \ref{tab:lranges}.}
\label{fig:harmplot}
\end{center}
\end{figure}

\begin{center}
\begin{table}
\caption{Multipole ranges\label{tab:lranges}}
\begin{tabular}{|c|c|c|c|c|c|c|c|c|c|c|c|c|c|c|}
\hline
bin  & 0 & 1 & 2 & 3 & 4 & 5 & 6 & 7 & 8 & 9 & 10 & 11 & 12  \\
$\ell_{min}$ & 2 & 2 & 2 & 2 & 2 & 7 & 7 & 7 & 7 & 7 & 10 & 20 & 30 \\
$\ell_{max}$ & 10 & 20 & 30 & 40 & 63 & 10 & 20 & 30 & 40 & 63 & 20 & 30 & 40 \\
\hline
\end{tabular}
\end{table}
\end{center}

In the previous section we showed that taking the whole map into account, we do not find any significant cross-correlation between the CMB and the predicted SZ effect. If some multipole ranges are contaminated by the Galaxy or if for some reason the SZ prediction is too inaccurate at certain scales, the pixel space test could give a negative result even if a cross-correlation can be seen at some multipoles. In this section, we have check the possibility of detecting the SZ effect in certain limited multipole ranges. In figure \ref{fig:harmplot} (left plot) we show the distribution of the estimates $\hat C$ for 1000 simulations of CMB in the 12 different multipole ranges indicated in table~\ref{tab:lranges}, the grey bands correspond to the 1 and 2 $\sigma$ levels. The crosses show the data values. Note that for some of the multipole ranges, we have excluded the first 5 multipoles which could have galactic residuals present (see i.e. Schwarz et al. 2004, Hansen et al. 2004). The results of a similar test on the difference map is shown in the same figure (right plot). 

As the CMB is dominating the variance on the largest scales whereas the noise dominates the variance on smaller scales, the  map 'contaminated' with CMB (left plot) has larger variance than the difference map (right plot) for the large scales. For the smaller scales, we see that the optimal map containing CMB does better. As in the previous section, we do not find any significant detection in any multipole range, but note further that the error bars on some ranges are too large to make any conclusion about the cross-correlation.

\subsubsection{Cross correlation in wavelet space}

\begin{figure}
\begin{center}
\leavevmode
\epsfig {file=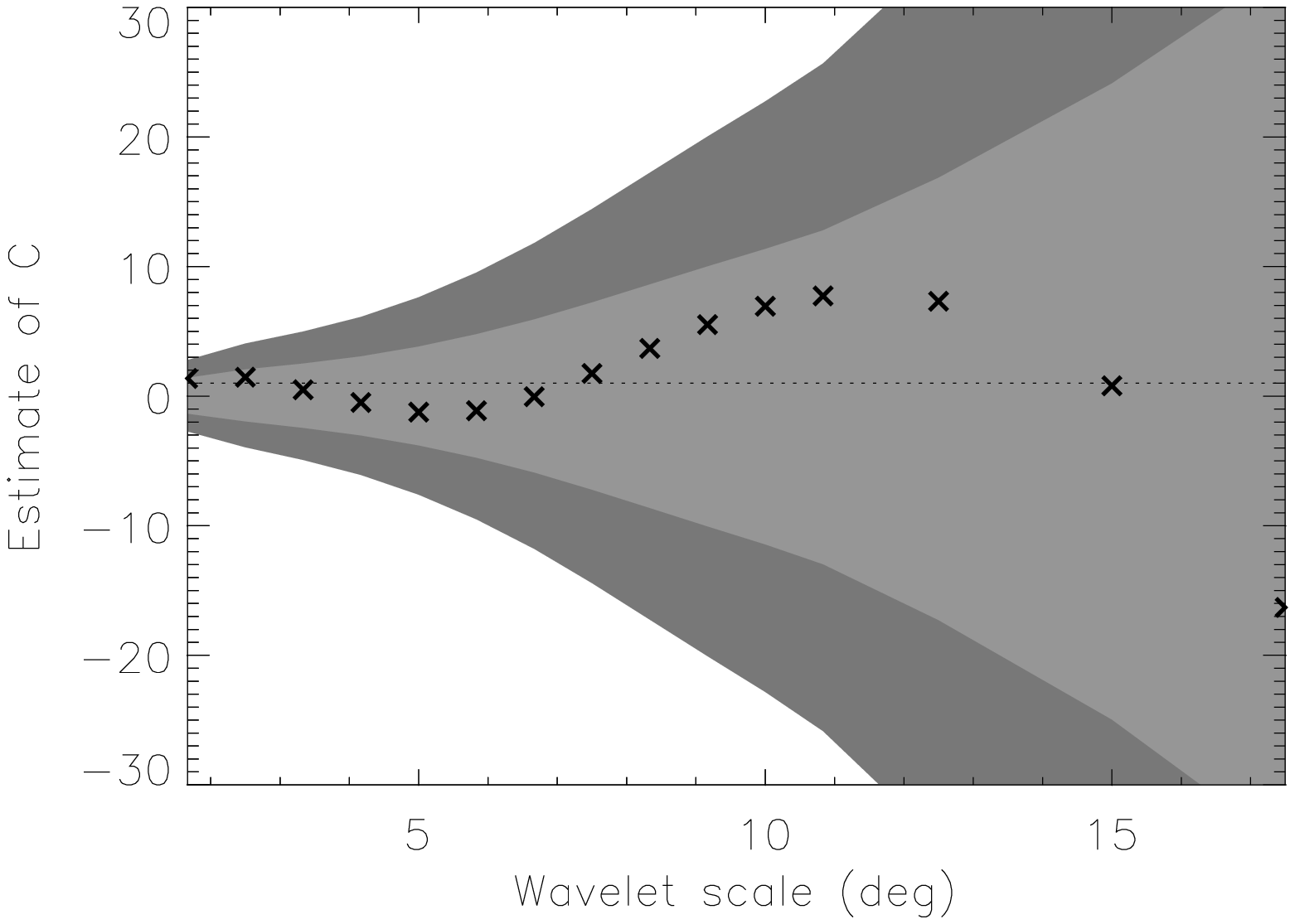,height=8cm,width=8cm}
\epsfig {file=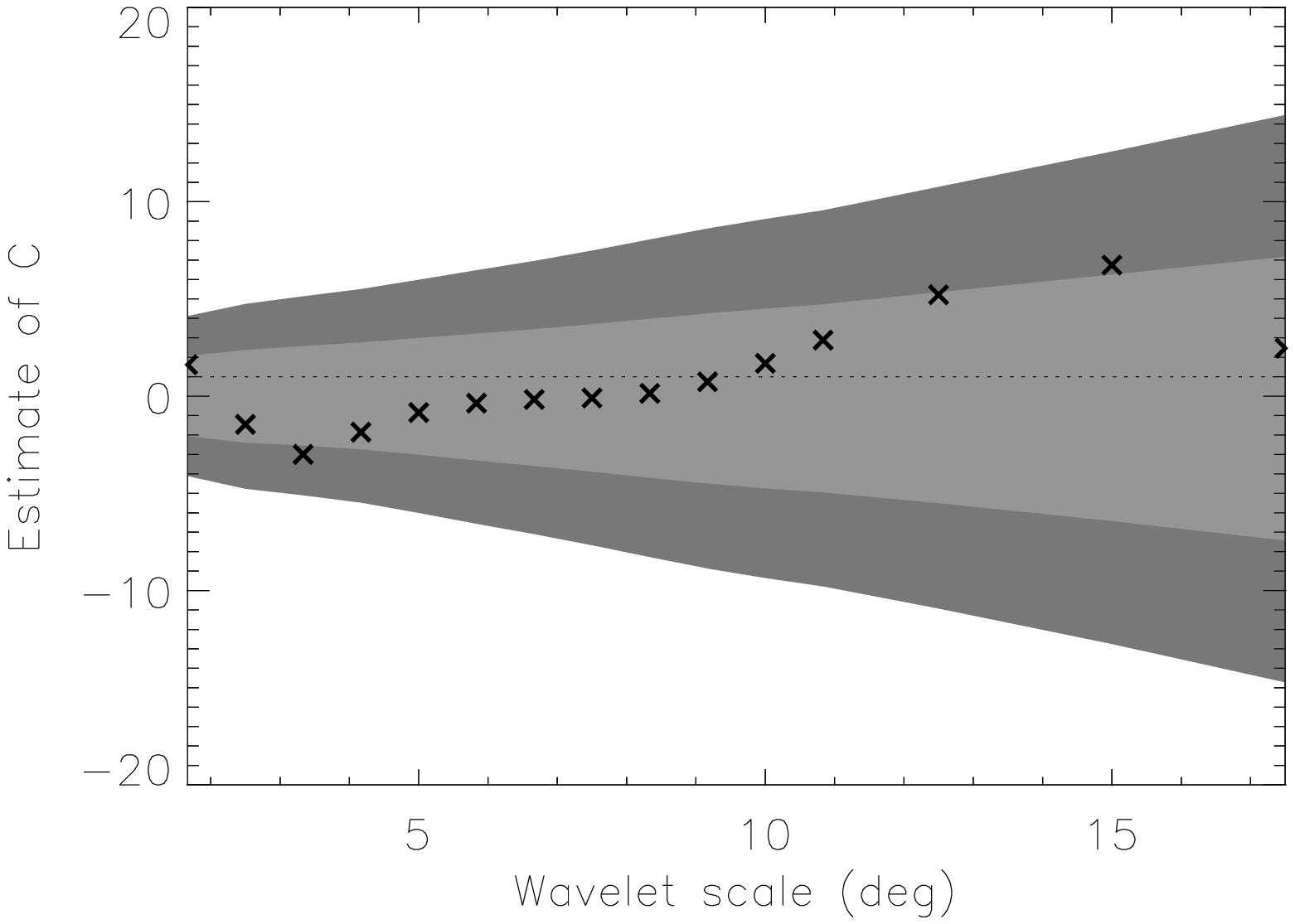,height=8cm,width=8cm}
\caption{The estimate of $C$ in wavelet space using the co-added V+W map (left plot) and the difference map (right plot). The grey bands show the 1 and 2 $\sigma$ level set from 1000 Monte-Carlo simulations of pure CMB maps, the dotted line shows the average of 1000 simulated CMB maps with the expected level of SZ-effect added. The crosses show the result from cross-correlation with the WMAP data.}
\label{fig:wavplot}
\end{center}
\end{figure}

Fig \ref{fig:wavplot} is analogous to fig \ref{fig:harmplot} for different wavelet scales. We see that wavelets give much stronger limits on the constant $C$ at a single scale than does the harmonic space test, but still we do not find any detection at any scale. Also for the difference map, we can only put upper limits. Combining all the scales as described in section \ref{sect:wavmeth} gives $C=0.87\pm1.3$ for the combined map and $C=0.17\pm1.69$ for the difference map, all at $1\sigma$, in agreement with the pixel space analysis. This was expected as the results from cross-correlation tests in different spaces are highly correlated when all information is included. Note again that the for the larger scales, the map containing CMB has larger variance whereas for the smaller scales the difference map containing only noise has largest variance.

\section{Conclusion}

\label{sect:concl}

We have used the galaxy density field obtained from the PSC$z$ galaxy survey and hydrodynamical simulations based on the observed
distribution of IRAS 1.2Jy galaxies to make a full-sky prediction of the SZ effect from diffuse hot gas in the local universe ($<80 \hmpc$). We have studied whether this effect could be observed and studied by the CMB observations performed by the WMAP satellite and if it could be a contaminating factor in the study of the CMB at large angular scales.

We have cross-correlated our map of the predicted local SZ-effect with the WMAP data at different frequencies taking into account the frequency dependence of the SZ effect. We have performed the cross-correlation test in three spaces, enabling us to test certain parts of the sky (pixel space test), certain multipole ranges (harmonic space test) and certain wavelet scales (wavelet test). Applying the pixel based cross-correlation test, we checked the full sky including all pixels outside of an extended galactic cut as well as smaller areas of the sky where the SZ effect was expected at a higher level. None of these tests gave a significant detection. The harmonic space test has the advantage that a cross-correlation occurring only at certain scales can be detected, but no detection was found for any multipole range. Finally, the wavelet transform is very efficient at amplifying structures at a specific scale and the cross-correlation test in wavelet space could reveal a possible correlation between the CMB and the SZ-effect from local hot gas, but no correlation was found at any scale. We found that the different tests have a very similar power to detect a possible cross-correlation when all the information is added and the three tests are strongly correlated as the same information is used in three different spaces. But as the map of the SZ effect could be inaccurate at certain scales or certain parts of the sky due to limited knowledge of temperature and gas densities, a test of cross-correlation at different scales or at different spatial positions is necessary. 

In general, we find the predicted effect too small to expect a clear detection using the 1 year WMAP data and only an upper limit can be set: at the $2\sigma$  confidence level, the SZ effect from diffuse hot gas in the local universe cannot exceed our predictions by more than 63\%. We find that the power specrum of our SZ prediction in the WMAP channels is 3-4 orders of magnitude lower than the power spectrum of the CMB. The local SZ effect is hence far too small to influence the estimate of the cosmological parameters from the WMAP data.

In this paper we have combined constrained simulations with data from the PSC$z$ galaxy catalogue to obtain the temperature and gas densities. One could also obtain maps of gas in the local universe from constrained simulations directly. The problem with this approach is that the structures are slightly shifted with respect to their real positions. For this reason, such maps are not very useful for cross-correlation studies, but they can be utilised for studying general statistical properties of the SZ effect from the local universe \cite{mauro}. This was also studied using random Hubble volume simulations by \cite{schaefer} where they were able to go deeper and to smaller scales than in the work presented here.

We also note that the WMAP frequency range is very limited. The Planck satellite experiment, scheduled for launch in 2007, will provide high resolution maps of the CMB in the frequency range from $30\textrm{GHz}$ to $857\textrm{GHz}$. The SZ effect is zero at $217\textrm{GHz}$, negative below and positive above. By taking advantage of a set of frequencies below and above the zero point, we expect to greatly enhance the detection power. Further, some of the Planck channels are expected to have a considerably lower noise level than the WMAP data which will improve the possibilities for seeing small signals such as the SZ effect from gas in the local universe. Finally, the huge range of frequency channels planned for the Planck experiment will allow subtraction of galactic foregrounds with a much higher precision. In this way, the risks for false detections due to foreground residuals is greatly reduced. Also the SZ map prediction can be improved by merging all sky galaxy redshift catalogues like PCS$z$ used here with all sky cluster catalogues like the Abell/ACO one. Not only CMB data will be improved in the future, also the galaxy surveys will go deeper with a high resolution. Including all these effects, we expect that CMB data could even be used for studying the properties of gas in the local universe. This will be investigated in a future paper.

\section*{Acknowledgements}
We are thankful to Saleem Zaroubi and Lauro Moscardini for useful discussions. FKH acknowledges financial support from the CMBNET Research Training Network. PM acknowledges support by European contract MERG-CT-2004-510143 and CXC grant G04-5155X. We acknowledge use of the HEALPix \cite{healpix} software and analysis package and the CMBfast software package \cite{cmbfast} for deriving the results in this paper. We acknowledge the use of the Legacy Archive for Microwave Background Data Analysis (LAMBDA). Support for LAMBDA is provided by the NASA Office of Space Science.

\end{document}